\documentclass[prd,superscriptaddress,showpacs,showkeys]{revtex4}
\usepackage[T1]{fontenc}
\usepackage{amsmath}
\usepackage{amssymb}

\usepackage{tabstackengine}
\usepackage{graphicx}
\usepackage[english]{babel}
\usepackage{amsfonts}

\usepackage{eurosym}
\usepackage{calrsfs}
\usepackage[usenames,dvipsnames,svgnames,table]{xcolor}

\begin{document}

\title{$P-v$ criticality of a specific black hole in $f(R)$ gravity
coupled with Yang-Mills field}

\author{Ali \"{O}vg\"{u}n}
\email{ali.ovgun@pucv.cl}

\affiliation{Instituto de F\'{\i}sica, Pontificia Universidad Cat\'olica de
Valpara\'{\i}so, Casilla 4950, Valpara\'{\i}so, Chile}

\affiliation{Physics Department, Arts and Sciences Faculty, Eastern Mediterranean University, Famagusta, North Cyprus via Mersin 10, Turkey}
\affiliation{TH Division, Physics Department, CERN, CH-1211 Geneva 23, Switzerland }

\begin{abstract}
In this paper, we study the $P-v$ criticality of a specific charged AdS type black hole (SBH) in $f(R)$ gravity coupled with Yang-Mills field. In the extended phase space, we treat the cosmological constant as a thermodynamic pressure. After we study the various thermodynamical quantities, we show that the thermodynamic properties of the SBH behave as a Van der Waals liquid-gas system at the critical points and there is a first order phase transition between small-large SBH. 
\end{abstract}


\date{\today}

\maketitle

\section{Introduction}

Important contribution on black holes's thermodynamics in anti-de-Sitter (AdS) spacetime is made by Hawking and Page \cite{hawking} where a first order phase transition is discovered between the Schwarzschild-anti-de-Sitter (SAdS) black holes that is known as the Hawking-Page transition. Then  Chamblin et al. and Cvetic  et al. show that the first order phase transition among Reissner Nordstrom (RN) AdS black holes and the similarities between charged AdS black holes as well as liquid gas systems in grand canonical ensemble \cite{Chamblin:1999tk,Chamblin:1999hg,Cvetic:2010jb}. Moreover, in the seminal papers of Kubiznak and Mann \cite{Kubiznak:2012wp}, the cosmological constant $\Lambda$ is used as dynamical pressure \cite{Dolan:2010ha} \begin{equation} P=-\frac{\Lambda }{8 \pi} = \frac{3}{8 \pi l^2} \end{equation} 
for the RN-AdS black holes in the extended phase space, instead of treating the $\Lambda$ as a fixed parameter (in standard thermodynamic) and its conjugate variable has dimension of volume 
\begin{equation} V=\left(\frac{\partial M}{\partial P}\right)_{S,Q}. \end{equation}
 Calculating the critical components and finding the phase transition of the RN-AdS black holes, it is shown that RN-AdS black holes behave similar to the Van der Waals fluid in the extended phase space where a first-order small/large black hole's phase transition occurs at a critical temperature below \cite{Gunasekaran:2012dq,Kubiznak:2014zwa,Kubiznak:2016qmn,Hennigar:2016xwd,Frassino:2016vww}. The Van der Waals equation 
 \begin{equation} (P+\frac{a}{v^2})(v-b)=kT, \end{equation}
 
 where its pressure is $P$, its temperature is $T$, its specific volume is $v=V/N$, the Boltzmann constant is $k$ and the positive constants are $a$ and $b$, takes into account the attractive and repulsive forces between molecules and gives an improved model for ideal gas behaviour to describe the basic properties of the liquid-gas phase transition with the ratio of $\frac{P_{c} V_{c}}{T_{c}}=\frac{3}{8}$ at critical points \cite{Hennigar:2016ekz,Hendi:2017fxp,Hansen:2016ayo,Mann:2016trh,Hennigar:2015esa,Delsate:2014zma}. Afterwards, applications of the thermodynamical law's to the black hole's physics have gain attention. Different researches are done by using the variation of the first law of thermodynamics of black holes and also application of the $P-v$ criticality on black holes \cite{Chen:2013ce,Cai:2013qga,Belhaj:2012bg,Bhattacharya:2017nru,Jafarzade:2017sur,Sadeghi:2015out,Liang:2017vac,Pradhan:2016feg,Zeng:2016fsb,Mo:2016ndm,Majhi:2016txt,Fan:2016rih,Miao:2016ipk,Zeng:2016aly,Mandal:2016anc,Chen:2016gzz,Dehyadegari:2016nkd,Hendi:2016njy,  Ma:2016aat,Liu:2016uyd,Poshteh:2016rwc,Guo:2016eie,Li:2016zca,Ma:2017pap,Upadhyay:2017fiw,Fernando:2016qhq,Fernando:2016sps,Sadeghi:2016dvc,Miao:2016ulg,Cheng:2016bpx,Guo:2016iqn,Liang:2016xrz,Zhang:2016yek,Zhang:2016yek,Mo:2016jqd,Dehghani:2016wmw,Zeng:2015wtt,Maity:2015ida,Hendi:2015hgg,Karch:2015rpa,Wei:2015ana,Xu:2015rfa,wenliu,Hendi:2015kza,Sherkatghanad:2014hda,Zhang:2014eap,Dehghani:2014caa,Hendi:2014kha,za1,za2,za3,za4,za5,za6,za7,za8,Zhao:2014raa,Li:2014ixn,Mo:2014qsa,Zhao:2013oza,Kuang:2016caz,Kuang:2017bst,Luo:2016zdo,Aranguiz:2015voa,Kastor:2009wy,Azreg-Ainou:2014twa,Azreg-Ainou:2014lua,Caceres:2015vsa,Johnson:2014yja,Momennia:2017hsc,Hendi:2018sbe,Hendi:2017uly}. Furthermore, AdS-CFT correspondence is the other reason for studying the AdS black hole. 
 
 In this paper we use a black hole's solution in the Yang-Mills field which is the one of the most interesting non-abelian gauge theory. By using the string theory models they find the Yang-Mills fields equations in low energy limit and then Yasskin found the first black hole solution in the theory of Yang-Mills coupled to Einstein theory \cite{yas}.
 
  Our main aim is to check $P-V$ criticality of a specific charged AdS type black hole (SBH) in $f(R)$ gravity coupled with Yang-Mills field (YMF) \cite{halilsoy} by comparing its result with the Van der Waals system. The Yang-Mills field is acted inside the nuclei with short range and $f(R)$ gravity which is an extension of Einstein's General Relativity with the arbitrary function of Ricci scalar $f(R)$ \cite{Chakraborty:2014xla,Nojiri:2013su,Nojiri:2014jqa,Nojiri:2017kex}. It would be of interest to study the $P-v$ criticality of SBH in $f(R)$ gravity coupled with YMF in the extended phase space treating the cosmological constant as a thermodynamic pressure. In this paper, we first study the thermodynamics in the extended phase space and then we obtain its critical exponents to show the existence of the Van der Waals like small-large black hole phase transitions.
  
  The paper is organized as follows: in Sec. II we will briefly review the SBH in $f(R)$ gravity coupled with YMF. In Sec. III we $P-v$ criticality of the SBH in $f(R)$ gravity coupled with YMF will be studied in the extended phase space by calculating its critical exponents. In Sec. IV we conclude with final remarks.

\section{SBH in $f(R)$ gravity coupled with YMF}

In this section, we briefly present a solution of SBH in $f(R)$ gravity
coupled with YMF with a cosmological constant in d-dimensions \cite{halilsoy}.
Then we discuss its temperature, entropy and other thermodynamic quantities.
The action of the $f(R)$ gravity minimally coupled with YMF ($c=G=1$
) is \cite{halilsoy}
\begin{equation}
S=\int d^{d}x\sqrt{-g}\left[\frac{f\left(R\right)}{16\pi}+\mathcal{L}\left(F\right)\right]
\end{equation}
where $f(R)$ is a function of the Ricci scalar $R$ and $L\left(F\right)$
stands for the lagrangian of the nonlinear YMF with $F=\frac{1}{4}tr\left(F_{\mu\nu}^{\left(a\right)}F^{\left(a\right)\mu\nu}\right)$
where the 2-form components of the YMF are $\mathbf{F}^{\left(a\right)}=\frac{1}{2}F_{\mu\nu}^{\left(a\right)}dx^{\mu}\wedge dx^{\nu}$.
Here the internal index $(a)$ for the degrees of freedom of the non-abelian
YMF. It is noted that this nonlinear YMF can reduce to linear YM field
($\mathcal{L}\left(F\right)=-\frac{1}{4\pi}F^{s}$ ) for $s=1$ and 
$f_{R}=\frac{df\left( R\right) }{dR}=\eta r$
which $\eta $ is a integration constant. Solving Einstein field equations for
the $f(R)$ gravity coupled with YMF give to the spherically symmetric
black hole metrics (Eq. 36 in Ref. \cite{halilsoy}) 

\begin{equation}
ds^{2}=-f(r)dt^{2}+\frac{dr^{2}}{f(r)}+r^{2}\left(d\theta _{1}^{2}+\Sigma_{i=2}^{d-2}
\Pi^{i-1}_{j=1} \sin ^{2}\theta _{j}\;d\theta _{i}^{2}\right)
\end{equation}
with $0\leq \theta _{d-2}\leq 2\pi ,0\leq \theta _{i}\leq \pi ,\text{ \ \ }1\leq
i\leq d-3$ in which the metric function $f(r)$ is

\begin{equation}
f(r)=\frac{d-3}{d-2}-\Lambda r^{2}-\frac{m}{r^{d-2}}-\frac{\left( d-1\right)
\left( d-2\right) ^{\frac{d-1}{2}}\left( d-3\right) ^{\frac{d-1}{4}}}{2^{%
\frac{d-5}{2}}\eta d}\frac{Q^{\frac{d-1}{2}}\ln r}{r^{d-2}}.
\end{equation}

Note that $\text{ }\varLambda=-\frac{1}{l^{2}}$, $M$ is the mass
of the black hole and $\eta$ is a constant.Furthermore for the limit of $Q^{7/2}\rightarrow0$,
it becomes well-known solutions in $f\left(R\right)$ gravity. The
Bekenstein-Hawking temperature \cite{ao1,ao2,ao3,ao4} of the black hole is calculated by $T$=$\frac{1}{4\pi}\frac{\partial f(r)}{dr}\mid_{r=r_{h}}$
\begin{eqnarray}
T= \,{\frac { \left( -1+ \left( d-2 \right) \ln  \left( r \right) 
 \right) {Q}^{d/2-1/2} \left( d-1 \right)  \left( d-2 \right) ^{d/2-1/
2} \left( d-3 \right) ^{d/4-1/4}+d{2}^{d/2-5/2} \left( -2\,{r}^{2}
\Lambda\,{r}^{d-2}+m \left( d-2 \right)  \right) \eta}{4\pi\,{r}^{d-2}
\eta\,d{2}^{d/2-5/2}r}}
,\label{t}
\end{eqnarray}
 where $r_{h}$ is the horizon of the black hole. and solving the
equation $f(r_{h})=0,$ the total mass of the black hole is obtained
as 
\begin{equation}
m=-{\frac {4\,{Q}^{d/2-1/2} \left( d/2-1 \right) ^{d/2}\ln  \left( r
 \right) \sqrt {2} \left( d-1 \right)  \left( d-2 \right)  \left( d-3
 \right) ^{d/4-1/4}+ \left(  \left(  \left( f-1 \right) d-2\,f+3
 \right) {r}^{d-2}+{r}^{d}\Lambda\, \left( d-2 \right)  \right) d
\sqrt {d-2}\eta}{\eta\,d \left( d-2 \right) ^{3/2}}}
\end{equation}
The entropy of the black hole can be derived as 
\begin{equation}
S=\frac{A_{h}}{4}\eta r_{h},
\end{equation}
where $A_{h}=\frac{d-1}{\Gamma \left( \frac{d+1}{2}\right) }\pi ^{\frac{%
d-1}{2}}r_{h}^{d-2}$ is the area of the black hole's event horizon. Then,
in the extended phase space, we calculate the pressure in terms of
cosmological constant 
\begin{equation}
P=-\frac{\varLambda}{8\pi} \label{cosmo}
\end{equation}
and its thermodynamic volume is

\begin{equation}
V=\frac{\Omega_{d-2} r_{h}^{d-1} \eta}{n-1},
\end{equation}
where $\Omega_{d-2}$ is the volume of the unit sphere. Now the mass can be also written in terms of $P$ as follows:
\begin{equation}
m=-{\frac {4\,{Q}^{d/2-1/2} \left( d/2-1 \right) ^{d/2}\ln  \left( r
 \right) \sqrt {2} \left( d-1 \right)  \left( d-2 \right)  \left( d-3
 \right) ^{d/4-1/4}+ \left(  \left(  \left( f-1 \right) d-2\,f+3
 \right) {r}^{d-2}-8\,{r}^{d}P\pi\, \left( d-2 \right)  \right) d
\sqrt {d-2}\eta}{ \left( d-2 \right) ^{3/2}\eta\,d}}.
\end{equation}

The first law of the black hole thermodynamics in the extended phase
space is 
\begin{equation}
dm=TdS+\varPhi dQ+VdP,
\end{equation}

where the thermodynamic variables can be obtained as $T=\left(\frac{\partial m}{\partial S}\right)_{Q,P}$
, $\varPhi=\left(\frac{\partial m}{\partial Q}\right)_{S,P}$ and
$V=\left(\frac{\partial m}{\partial P}\right)_{S,Q}$.

Then we write the generalized Smarr relation for the black hole, which
can be derived also using the dimensional scaling, as

\begin{equation}
m=2TS+\varPhi Q-2VP.
\end{equation}

We introduce the cosmological constant as thermodynamic pressure in
the extended phase space in Eq. (\ref{cosmo}), and it is seen that the
first law of the black hole's thermodynamics and the Smarr relations are
matched well. 

\section{$P-v$ Criticality}

In this section, we investigate the critical behaviour of the SBH
in the extended phase space. The critical point can be defined as 
\begin{equation}
\frac{\partial P}{\partial v}=\frac{\partial^{2}P}{\partial v^{2}}=0. \label{crt}
\end{equation}
Now we consider the case of four dimensions $(d=4)$ where the metric function becomes 
\begin{equation}
	f=\frac{1}{2}-\frac{m}{r^2}-r^2 \Lambda - \frac{3 Q^{3/2} ln(r)}{{r^2} \eta}
\end{equation} and corresponding mass of the black hole is calculated as 

\begin{equation}
m=\frac{r^2}{2}-r^2 \Lambda	-\frac{3 Q^{3/2} ln(r)}{\eta}.
\end{equation}

The temperature of the four dimensional SBH is \begin{equation}
	T=-\frac{3Q^{3/2}-r^2\eta+4r^4 \eta \Lambda}{4 \pi r^3 \eta}. 
\end{equation}

Then we write the temperature in terms of P ($P=-\frac{\varLambda}{8\pi}$) as follows:

\begin{equation}
T=8\,rP-\,{\frac {3{Q}^{3/2}}{4{r}^{3}\eta\,\pi}}+\,{\frac {1}{4r\pi}
}
. \end{equation}
 Afterwards one can easily obtain the pressure P in terms of the temperetaure T: 
 
 \begin{equation}
 	P=\,{\frac {T}{8r}}+{\frac {3\,{Q}^{3/2}}{32\,{r}^{4}\eta\,\pi}}-
\,{\frac {1}{32{r}^{2}\pi}}
.
 \end{equation}

To consider the $P-v$ criticality using the extended phase space,
we write the black hole radius in terms of the specific volume $v$
as $ r_{h}=\frac{(d-2) v}{4}$. Using the condition of Eq.(\ref{crt}), we derive the critical Bekenstein-Hawking
temperature $T_{c}$, critical pressure $P_{c}$ and critical specific
volume $v_{c}$ as follows:
\begin{equation}
T_{c}=-24\,{\frac {{Q}^{3/2}}{{v}^{3}\eta\,\pi}}+{\frac {1}{v\pi}}
,
\end{equation}

\begin{equation}
v_{c}=24\,{\frac {\sqrt {2}\sqrt {\,{Q}^{3/2}}}{\sqrt{\eta}}},
\end{equation}

\begin{equation}
P_{c}={\frac {\eta}{1152\,{Q}^{3/2}\pi}}.
\end{equation}

One can also find this relation which is same with a Van der Waals
fluid
\begin{equation}
\rho_{c}=\frac{P_{c}v_{c}}{T_{c}}=\frac{3}{8}.
\end{equation}

It is noted that in Fig.1 and Fig.2 show that $P-r$ diagram is same with the
diagram of the Van der Waals liquid-gas system.

\begin{figure}
\centering
\includegraphics[width=4in]{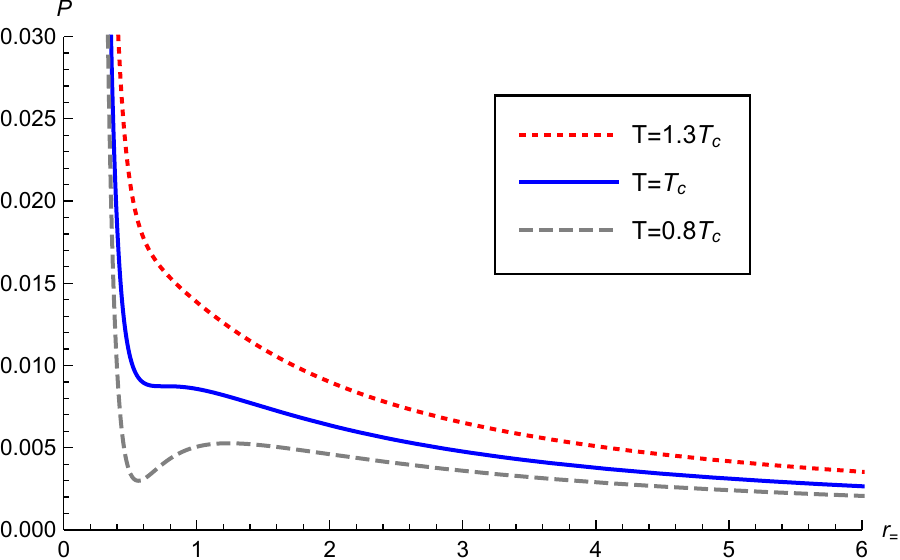}
%\vspace{-8cm}

\caption{$P-r$ diagram of a SBH in a $f(R)$ gravity
coupled with YMF for $Q=0.1$ and $\eta=1$}

\label{Fig.1}
	
\end{figure}

\begin{figure}
\centering
\includegraphics[width=4in]{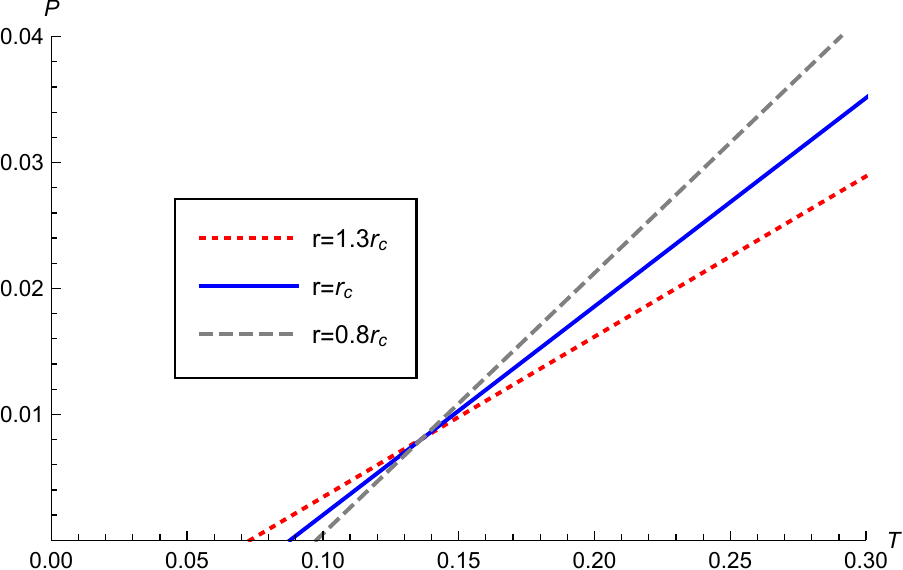}
%\vspace{-8cm}

\caption{$P-T$ diagram of a SBH in a $f(R)$ gravity
coupled with YMF for $Q=0.1$ and $\eta=1$}

\label{Fig.2}
	
\end{figure}

Let us now analyze the Gibbs free energy of the system. We first use the mass as entalphy instead of internal energy and the Gibbs free energy in the extended phase space for the SBH in $f(R)$ gravity coupled with Yang-Mills field is calculated as
\begin{equation}
G=m-TS=\frac{6 Q^{3/2}+r^2\eta -16 P \pi r^4 \eta -18 Q^{3/2}ln(r)}{6 \eta}.
\end{equation}

	\begin{figure}
\centering
\includegraphics[width=4.0in]{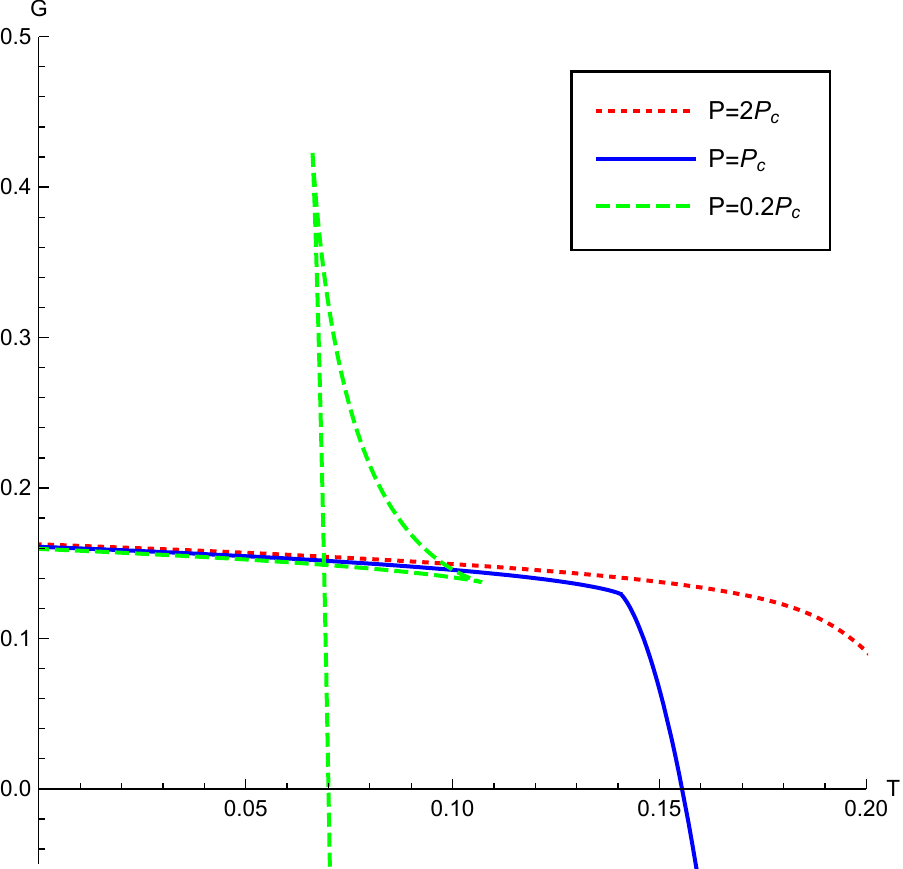}
%\vspace{-6.0cm}
\caption{$G-T$ diagram of a SBH in a $f(R)$ gravity
coupled with YMF for different values of P ( $P<P_{c}$, $P=P_{c}$ and $P>P_{c}$ with $Q=0.1$ and $\eta=1$}
\label{Fig.3}
	\end{figure}

We plot the change of the free energy $G$ with $T$ in Fig.(3). There is a small-large black hole phase transition as seen in Fig. (3).

\section{Conclusion}

In this paper, we first treat the cosmological constant $\Lambda$ as a thermodynamical pressure $P$ and the thermodynamics and $P-v$ criticality  of the SBH in $f(R)$ gravity coupled with YMF is studied in the extended phase space. It is shown that 
 there is a phase transition between small-large black holes. Furthermore, after we obtain the critical exponents, the critical behaviour of SBH in $f(R)$ gravity coupled with YMF in the extended space behaves also similarly as Van der Waals liquid gas systems with the ratio of $\frac{P_{c} v_{c}}{T_{c}}=\frac{3}{8}$ at critical points. Hence it would be of great importance to obtain the $P-V$ criticality of SBH in $f(R)$ gravity coupled with YMF. Hence the critical ratio $\frac{P_{c} v_{c}}{T_{c}}=\frac{3}{8}$ is universal and independent from the modified gravities.
 The YMF has a parameter of $\eta$ but has no effect on the universal ratio of $\frac{3}{8}$.

 It is also interesting to study the holographic duality of SBH in $f(R)$ gravity coupled with YMF. It is noted that without thermal fluctuations black hole is holographic dual with Van der Waals fluid  given by 
$\left(P+\frac{a}{V^{2}}\right)\left(V-b\right)=T$ where k is the Boltzmann constant \cite{Caceres:2015vsa,Johnson:2014yja}, $b >0$ is nonzero constant which is the size of the molecules of fluid and the constant $a > 0$ is a value of the interaction measurement between molecules. We leave this problem for the future projects.

\acknowledgments
This work was supported by the Chilean FONDECYT Grant No. 3170035 (A\"{O}). A\"{O} is grateful to the CERN theory (CERN-TH) division, Waterloo University, Department of Physics and Astronomy, and also Perimeter Institute for Theoretical Physics for hosting him as a research visitor where part of this work was done. It is pleasure to thanks Prof. Robert B. Mann for valuable discussions. The authors declares that there is no conflict of interest regarding the publication of this paper.

\end{document}